\documentclass[12pt]{article}
\usepackage{a4}
\usepackage{graphicx}
\usepackage{amsfonts}
\usepackage{amscd}
\usepackage{latexsym}
\usepackage{epsf}
\usepackage{amsmath}
\numberwithin{equation}{section}

\newcommand{\be}{\begin{equation}}
\newcommand{\ee}{\end{equation}}

\newcommand{\bea}{\begin{eqnarray}}
\newcommand{\eea}{\end{eqnarray}}

\newcommand{\eqq}{equation }

\newcommand{\la}{ \lambda }

\newcommand{\om}{ \omega }

\begin{document}

\title{ How Long Is a Photon?}
\author{
{\sc I.~V.~Drozdov }\thanks{e-mail: drosdow@uni-koblenz.de} \\
{\small and}\\
A.~A.~Stahlhofen\thanks{e-mail: alfons@uni-koblenz.de}\\
\small  University  Koblenz\\
\small Department of Physics\\
\small  Universit\"atsstr.1, D-56070 Koblenz, Germany} \maketitle

\begin{abstract}
 An interpretation of an electromagnetic quantum as a single pulse is suggested. In this
 context the Planck formula is shown to be equivalent to the Heisenberg time-energy uncertainty
  relation.
 This allows to treat the photon frequency as an inverse time of emission.
 Such an ansatz avoids the inherent problems of the conventional approach.
\end{abstract}

\section{Introduction}

  A number of interesting developments in optical technologies
  in the past twenty years preset the "Century of the photon" in applied sciences and
  industrial technics. This concerns in particular the interaction of electromagnetic fields with matter
  in the subwavelength scale, i.e. in
 nano- and femto- regions of space and time. On this scale the photon is, however, not a point-like particle anymore, but a certain
 space-time structure of the electromagnetic field.

This raises the question, if the well known "macroscopic" assumptions on the structure and description of photons are justified on this scale.
 A (partial) answer to this question is presented here.

The famous ${\cal E}=h\nu$ of Max Planck (1900) was based on the
following at this time known facts \cite{mehra}:

 1. Energy is emitted from "atoms" in the form of electromagnetic waves.

 2. These waves were characterized as propagating continuous harmonic oscillations.
  On the basis of this model, these waves could be assigned corresponding
  frequencies $\nu$, which were then determined experimentally.

 3. The next step was the postulate of energy quantization in terms of the frequencies defined above:
 With the fixed frequency $\nu$ the total energy of the emitted electromagnetic wave is an integer
  multiple of $h\nu$.

 4. From these postulates originated the hypothesis of an elementary portion of the EM-field, thereby
paving the way for the concept of a "frequency" into the microscopic quantum domain.

Although the measurement of frequency, for instance, for
ultra-short pulses is highly sophisticated and very successful,
the purpose of this short note is to suggest an interpretation of
these data avoiding intrinsic inconsistencies.

\section{Frequency and photon structure}

 The energy of a single electromagnetic field portion is in general considered to be
given by the Planck formula
 \be {\cal E}=h \nu. \label{planck} \ee

Combining this formula with $c=\la\nu$, i.e. the propagation
velocity of the electromagnetic field, one obtains a corresponding
wavelength  $\la=h c/{\cal E}$.

 This leads already to an apparent inconsistency between the conventional treatment of a photon
  as a particle with a certain coordinate, energy, momentum and velocity and simultaneously the
  spatial delocalization of order $\la$. The declaration of the "wave-particle dualism" appointed to
  save the classical imagination of particle, does not abolish the question of the real space-time
  geometry of the sufficiently non-localized electromagnetic field.

\includegraphics[scale=0.34]{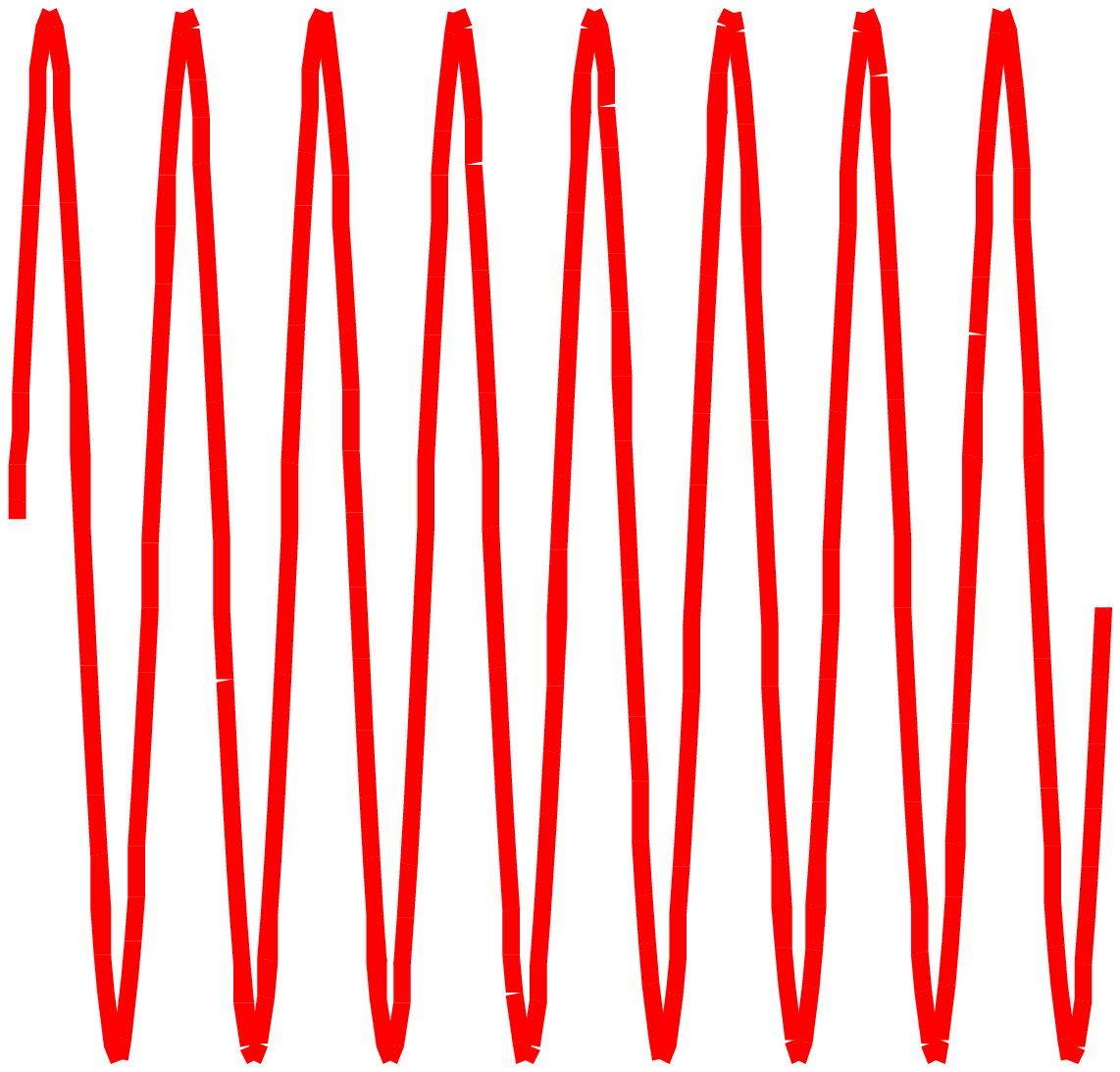}\includegraphics[scale=0.34]{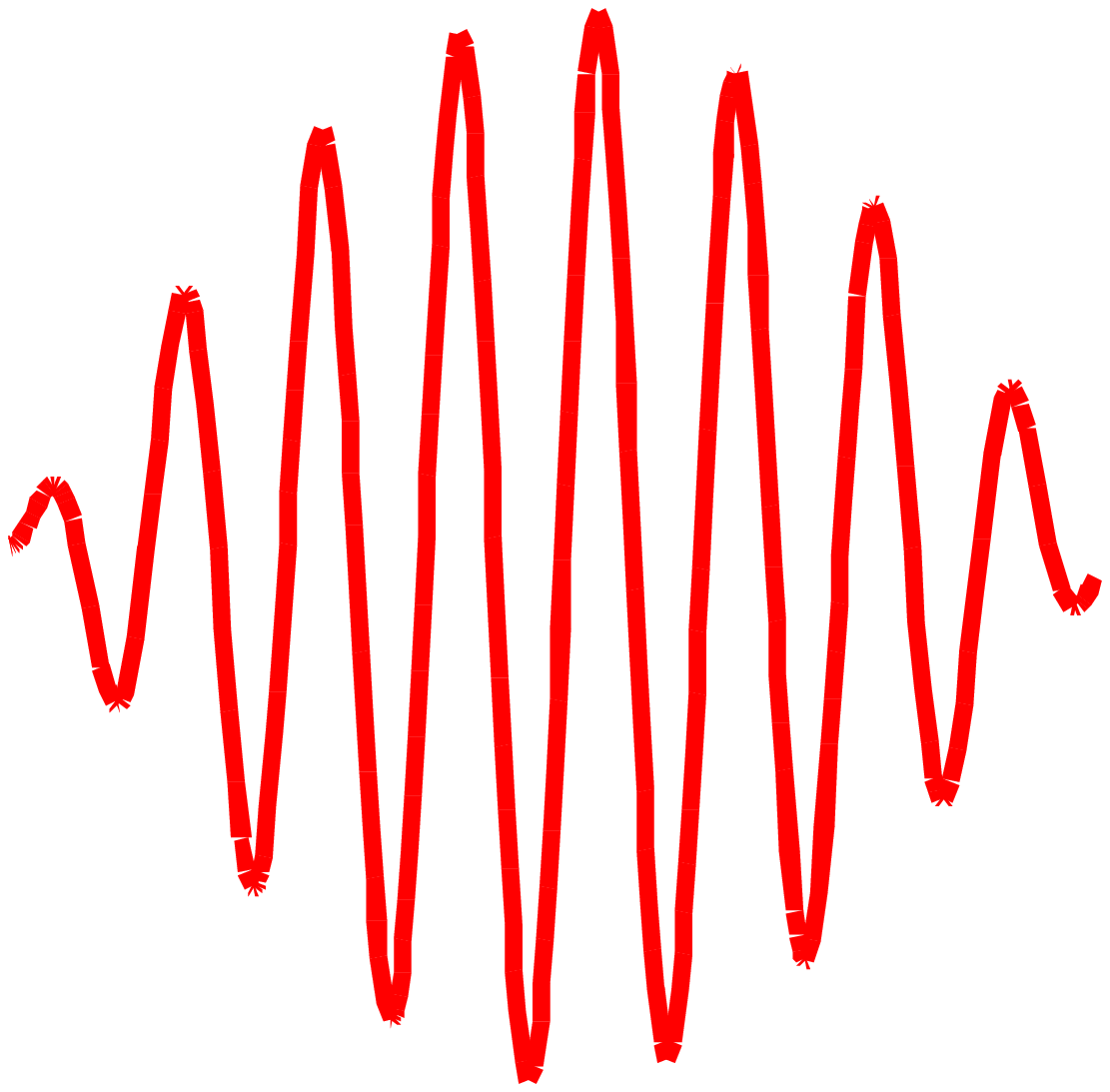}\includegraphics[scale=0.34]{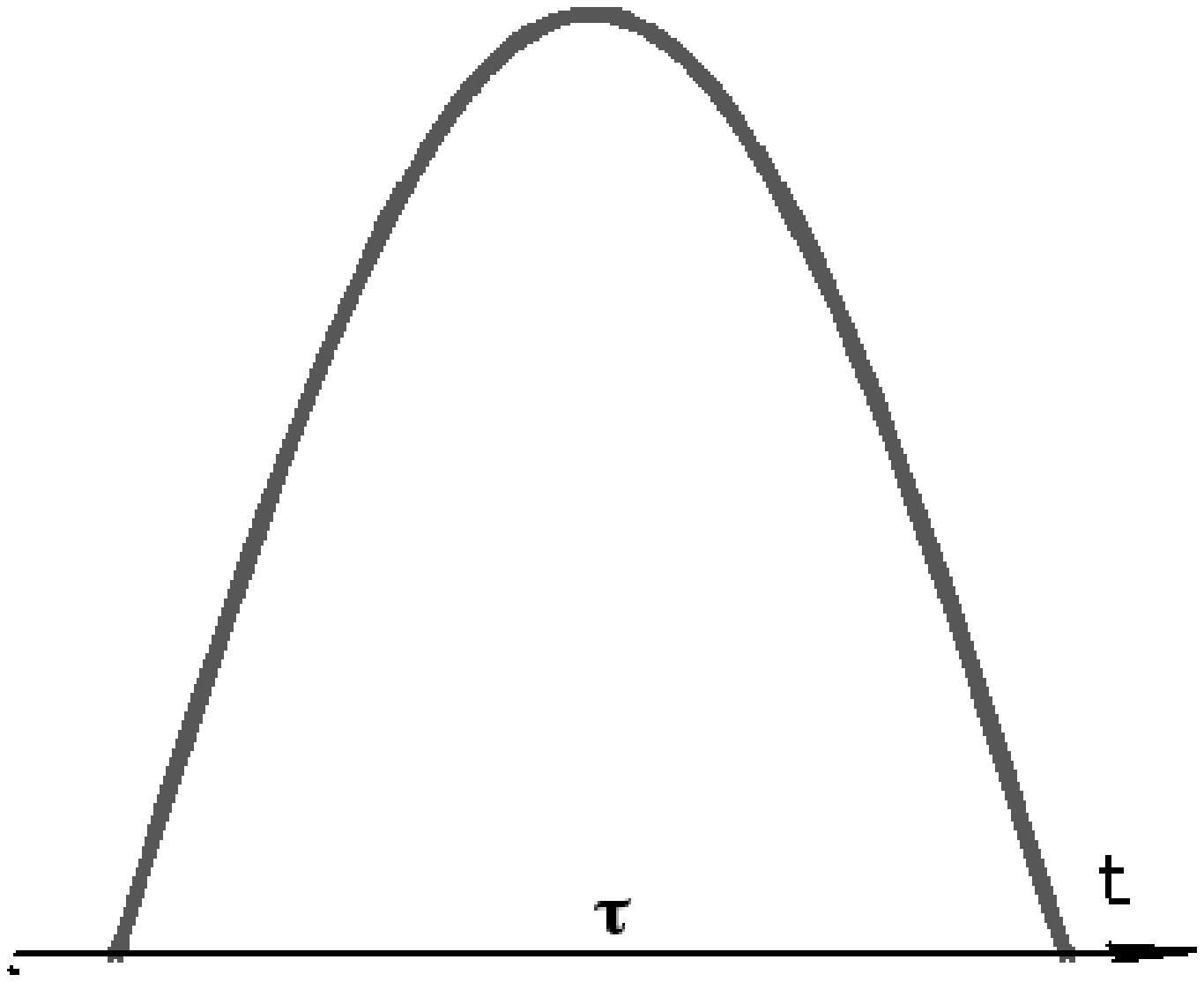}

{\bf\small\hspace*{2cm} Fig.1 \hspace*{3cm} Fig.2 \hspace*{3cm}  Fig.3 \hspace{2cm}\\}

A single photon is conventionally thought as a "wellenzug" (wave
train) - a cut of (sinusoidal) harmonic oscillations - like the
one shown in Fig.1

 This representation leads to the question: how long is this wellenzug, or
in terms of periods/frequency - how many periods of oscillations
does it contain?
 The complete "wellenzug" should than be embedded in a domain of its spatial localization
 with the corresponding frequency $\nu$ and the field amplitude $A_0$.
  The total energy of this field configuration is then proportional to the frequency and the
  squared amplitude, i.e. the intensity. The remaining factor should contain a (convergent) spatial integral of the field shape over the localization domain.

 It is immediately obvious, that the relation between the "length" of photon (or number of periods therein) and the
field amplitude is ambiguous and not fixed by the Planck relation.
Equation (\ref{planck}) is in fact a {\it definition} of the
frequency $\nu$, allocating this value of the frequency to an
equivalence class of field shapes with finite energy ${\cal E}$ .

 From a different point of view, the relation of the squared amplitude and the
global time duration of the emission process is restricted by the
Heisenberg uncertainty relation for time and energy,
  \be
  \Delta{\cal E}\tau \sim A_0^2\tau \sim h/2.
  \label{heisenberg}
\ee
This relation shows, that a shortest photonic time
corresponding to the photonic energy at hand is of order of the
half oscillation period.
 But applied to the Planck ansatz of a harmonic periodicity, the half-period
means indeed nothing  other else than a single pulse without any periodicity, treated
in quantum mechanics as a typical "wave packet" (Fig.3).

  From this point of view, there are no oscillations with the frequency $\nu$; the present approach takes as a
fundamental quantity the characteristic time (or process duration)
$\tau= T/2 = 1/2\nu$.
 In this context, the Planck formula (\ref{planck}) provides the same information as
Heisenberg's relation (\ref{heisenberg}). Such a very special
sample as the half-period-cut of a sinusoidal shape concerns in
applications the concept of so called "monochromatic" pulses. A
pulse having a non-sinusoidal shape is being called usually a
 "non-monochromatic" one respectively. To emphasize: the term
of (non-)monochromaticity of the single pulse solely concerns the
treatment in terms of Fourier series, which are
 nothing more than a conventional mathematical tool.


  The "good old Planck formula" becomes simply not applicable in these cases since
   the notion of "frequencies" for arbitrary single pulse is not well-defined as outlined
 above.

  For the same reason a representation of an arbitrary single pulse as a superposition of states
  with different energies (corresponding to multiplicity of frequencies) is
  a pure mathematical problem of representation and provides no physical
  meaning.
 For photon emission by a single atom, observed in a pure initial and a pure final state, the global amount of the emitted EM-energy is defined clearly and the global energy conservation
 remains valid also in microscopic scale.

 Generally, a geometry of an EM field is governed by Maxwell's field equations. In a quantum-mechanical interpretation, an arbitrary
  spatial field geometry must not necessarily be an eigenfunction of any Hamiltonian,
 but can be decomposed in the basis of eigenstates of this Hamiltonian (or another operator). It does not mean,
 the state with a given geometry configuration possesses "a wide spread of energy", since a Hamiltonian
 operates with spatial degrees of freedom, while energy is an invariant of time-translations. Therefore
 the "wide spread" is merely a problem of representation of an arbitrary state by a certain (in general, arbitrary constructed)
 Hamiltonian.

 On the other side, the equations of electrodynamics show, that the
shape and spatial localization/delocalization are not fixed by
them completely. In particular, the wave equation for the photon
field does not contain any information about the harmonic
oscillations and their frequency $\nu$ respectively.
 This information for the parameter $\nu$ has to be obtained therefore
from the emitting system. Otherwise an arbitrary shape $\psi$ of e.g. translation modes $\psi(t\pm x/c)$
 obeys the wave \eqq as well.

Since the emission of photons is expected in the form of harmonic
oscillations, the emitting system itself should at least in
principle be viewed as an oscillator. To check this conventional
assumption, let us consider a single atom as an emitting system.
The conventional description of classical optics treats a single
atom of an optical medium indeed as a harmonic oscillator.
 Despite all applications based on this assumption, it should be kept in mind that these
 conditions can never be achieved in reality.
 Any oscillator - especially a linear one - is determined by a special form of the potential,
that should possess a (local) minimum (vacuum) and allows a
quadratic approximation in the vicinity of the vacuum point.
 It is obvious, however, that already the Coulomb central potential of the atom never obeys these
conditions, since it does not allow in any point an expansion containing the required quadratic term.

This shows that the emission of a photon upon transition from an
excited state to a lower one does not reveal any oscillations and
should be characterized by a characteristic duration $\tau$ rather
than by frequency. The emitted photon is a single pulse of a
spatial length corresponding to this parameter as $\la\sim c\tau$
(Fig.3).

 A "refined" model of a photon (Fig.2) maintains the shape of a "wellenzug" like Fig.1, multiplied with an "enveloping" shape like the single pulse of Fig.3.
 Apart from the discussion of linear oscillations above, the following remark is in order: the shape Fig.2 contains
effectively two characteristic parameters with the dimension of
time - the general emission duration $T$ and the single
oscillation period $\tau=1/\nu$. While one of them can be related
to the total emitted energy via (\ref{heisenberg}), it is unclear,
what the second parameter is related to.

\section{Conclusion}

 A critical review of the well known concept of a photon frequency reveals inconsistencies in the conventional picture
 and interpretation of the photon structure.
 A treatment of the photon as a single pulse of the electromagnetic field is possible as argued above without raising contradictions
 with basic principles of quantum mechanics and electrodynamics. The frequency, defined by the Planck relation
 should be treated in this context as a characteristic time duration $\tau\sim 1/\nu$.
  The Planck formula itself signifies nothing about the photon intrinsic structure (especially oscillations) and concerns entirely a certain
  equivalence class of field configurations with respect to the total energy.

  Finally, this formula is in fact the same statement as Heisenberg's time-energy uncertainty relation.
 The origins of any intrinsic photonic oscillations, usually taken for granted, are not justified.
  Especially an emitting atom cannot provide them since the visualization of an atom as a harmonic oscillator is inconsistent.

  The canonical ansatz describing a single photon using harmonic oscillations of the electromagnetic field of type $e^{i\om t}$
  seems to work properly because the usual semiclassical approach is based on global (integral) expressions of
   this ansatz; the results so obtained are not related to them directly due to the inherent ambiguity of integration results,
   while containing the original frequency $\om$ as a parameter.

This is the interpretation of the experimental frequency data
mentioned above.

 Taking the classical oscillation ansatz for granted keeps an investigation of the
 photon wave function in the sense of the real geometry of the spatial electromagnetic field distribution shape of the
 field quantum, or at least an equivalence class of these shapes, still out of theoretical interest.

 It appears more appropriate to consider a single photon like a real ultra-short pulse;
 the frequency is then understood as an inverse of the emission time $\tau$
 instead of an ambiguous wave length. It is characterized by a well defined spatial localization, related to them
 as $\la\sim c\tau$. The term "frequency" could be entirely interpreted as a parameter of the leading component of
 the Fourier decomposition of the pulse shape, meaning that the frequency makes sense in a purely technical,
  but not a physical context.
For instance, Gaussian wave packets are perfect illustration of
short photon pulses. The Gaussian pulse is a certain shape
$e^{-t^2}$ (or $e^{-(t\pm x/c)^2}$ as a travelling wave), having a
certain amount of energy defined by its geometry as $\int T^{00}
d^3x$ (00-component of the energy-momentum tensor) of the EM-field
but having no frequencies.

   This concept itself does not contradict any of present theories; especially, a theory would not change
    due to replacing the term "frequency" by the "inverse time".
  Moreover, an existence of conventional intrinsic oscillations in the photon, emitted by a single atom,
 apparently contradicts the theoretical foundations (of classical and quantum mechanics), as pointed out
 above.

  An uncertainty of any experimental confirmation or
  contradictions beats on experiments where oscillations inside of single photon
   could have been observed directly.
  All indirect conclusions thereon are entirely a question of
 {\it interpretation} drawn {\it a priori} on the oscillation ansatz.

 The aim of the paper was finally to put attention to the
possibility to forego the interpretation in terms of frequencies.
Especially in the case of the single
 pulse it's more fruitful to investigate it "as it is" instead of its Fourier decomposition
 usually used, i.e. without any non-existent oscillations and its frequencies;
 the same concerns also photons.

 The single photon {\it must not} necessarily be a wave train containing intrinsic oscillations,
  but it is rather a single pulse and it {\it may be} viewed as such one.

\end{document}